\documentclass[twocolumn,showpacs,preprintnumbers,prb,superscriptaddress]{revtex4}
\usepackage{graphicx}
\usepackage{amssymb}
\usepackage{amsmath}

\begin{document}

\title{Flexuron, a self-trapped state of electron in crystalline membranes}
\author{M. I. Katsnelson}
\affiliation{Radboud University of Nijmegen, Institute for
Molecules and Materials, Heyendaalseweg 135, NL-6525 AJ Nijmegen,
The Netherlands}
\date{\today}

\begin{abstract}
Self-trapping of an electron due to its interaction with bending
fluctuations in a flexible crystalline membrane is considered. Due
to the dependence of the electron energy on the corrugations of
the membrane, the electron can create around itself an anomalously
flat (or anomalously corrugated, depending on the sign of the
interaction constant) region and be confined there. Using the
Feynman path integral approach, the autolocalization energy and
the size of the self-trapped state (flexuron) are estimated. It is
shown that typically the size of the flexuron is of the order of
the wavelength of fluctuations at the border between harmonic and
anharmonic regimes. The flexuron states are connected with the
fluctuation tail of the electron density-of-states, the asymptotic
behavior of this tail being determined by the exponent of the
renormalized bending rigidity.
\end{abstract}


\maketitle

\section{Introduction}

Studies of statistical mechanics in two dimensions \cite{nelson}
have been strongly stimulated recently by the discovery of
graphene \cite{first}, the first truly two-dimensional crystal and
the simplest possible membrane. Experimental \cite{jannik} and
theoretical \cite{Fas2007} demonstrations of intrinsic ripples on
graphene due to thermal bending fluctuations have initiated
numerous works on the effect of the ripples on the electronic
properties of graphene
\cite{KG08,castroneto,oppen,gazit,VKG,guinea} due to coupling
between electronic and lattice degrees of freedom. This coupling
originates from several factors, such as the dependence of
electron hopping parameters on interatomic distances and on the
angles between chemical bonds as well as from the redistribution
of electron density in a deformed membrane (see for review Refs.
\onlinecite{VKG,rmp}).

However, graphene is just the first representative of a huge class
of two-dimensional crystals including broad-gap semiconductors
such as hexagonal boron nitride \cite{pnas}, graphane
(hydrogenated graphene) \cite{graphane} and fluorographene
\cite{fluor1,fluor2}. Electrons in such materials are just
conventional nonrelativistic quantum particles and not chiral
Dirac fermions like in graphene, and one cannot expect there
manifestations of exotic effects of corrugations such as gauge
fields \cite{VKG}. At the same time, another interesting physics
specific of the two-dimensional case arises which will be the
subject of the present work.

It is known since many years that the interaction of a charge
carrier in a semiconductor with some order-parameter fluctuations
can drastically change its state leading to self-trapping, or
autolocalization
\cite{brinkman,krivoglaz,ourTMF1,ourJMMM,ourTMF2,nagaev,dagotto,AK1,AK2}.
This phenomenon is of crucial importance, for example, for the
phase separation in magnetic semiconductors and colossal
magnetoresistance materials \cite{ourTMF2,nagaev,dagotto}, where
the magnetization plays the role of order parameter. Since the
band motion of the electron is easier (and, hence, the bandwidth
is larger) for a ferromagnetically ordered state the electron in
antiferromagnetic or magnetically disordered surroundings creates
a ferromagnetic region and turns out to be self-trapped in this
region (different names are used for this state, such as spin
polaron \cite{ourJMMM}, ferron \cite{nagaev}, or fluctuon
\cite{krivoglaz}). If concentration of electrons is large enough
it leads to a formation of ferromagnetic state in the whole
crystal, via so called double exchange, otherwise the phase
separation happens, all electrons being trapped in the
ferromagnetic regions. The crucial mechanism is the dependence of
electron hopping parameter on the angle between magnetic moment on
neighboring sites \cite{brinkman,ourTMF2}. Importantly, the
critical point where fluctuations of the order parameter are the
strongest is the most favorable for these ``fluctuon'' effects
\cite{ourTMF1,ourJMMM,AK1,AK2}.

Since the hopping parameter in a fluctuating membrane depends on
the angle between normals \cite{castroneto} one can expect a
similar physics. If the electron motion is the easiest in the flat
membrane, self trapping in an anomalously flat region is possible,
in the opposite case an anomalously crumpled region can arise.
Since membranes are characterized by very strong bending
fluctuations and very slow, power-law decay of the corresponding
correlation functions \cite{nelson} the situation is formally
similar to that for autolocalization in the critical point in
three-dimensional systems. I will study here this autolocalized
state which can be called ``flexuron'', by analogy with
``ferron'', or ``fluctuon''. Technically, I will use the path
integral approach \cite{path1,path2,path3} as was applied to the
fluctuon problem in our previous works
\cite{ourTMF1,ourJMMM,AK1,AK2}. This method is an adaptation of
the seminal work by Feynman on the theory of polaron
\cite{feynman}.

It will be shown that the flexuron forms in crystalline membranes
at a large enough coupling constant and its size is of the order
of the length scale where interactions between bending and
stretching modes become important. The number of electron states
in the fluctuation tail of the electron density of states where
all states are (auto)localized, even without extrinsic disorder,
turns out to be proportional to the temperature.

\section{Description of the model and mean-field analysis}

In the continuum medium theory \cite{nelson} the membrane is
described by the out-of-plane-deformation field $h\left(
x,y\right) $. The unit normal vector at a given point is
\begin{equation}
\mathbf{n}\left( x,y\right) =\frac{\left( -\frac{\partial h}{\partial x},-%
\frac{\partial h}{\partial y},1\right) }{\sqrt{1+\left| \nabla
h\right| ^2}} \label{normal}
\end{equation}
where $\nabla $ is the two-dimensional gradient. In the ground
state, the membrane is supposed to be flat, and all normals are
along $z$ direction (Fig. 1a). At finite temperatures, the
membrane is fluctuating, and the normals are no more parallel
(Fig. 1b). Let us assume, for simplicity, that the electron
spectrum is isotropic near the band minimum. Then, by symmetry,
the shift of the band edge due to bending fluctuations should be
proportional to  $\left| \nabla h\right| ^2$ and the electron
Hamiltonian takes the form
\begin{eqnarray}
\mathcal{H}_e &=&-\frac{\hbar ^2}{2m}\nabla ^2+g\varphi \left( \mathbf{r}%
\right) ,  \nonumber \\
\varphi \left( \mathbf{r}\right)  &=&\left| \nabla h\left(
\mathbf{r}\right) \right| ^2  \label{hamm}
\end{eqnarray}
where $\mathbf{r}=(x,y)$, $m$ is the effective mass, and $g$ is
the coupling constant, of the order of the electron bandwidth. It
can be both positive or negative, as a result of the interplay of
different contributions. If the band is narrowed under crumpling,
$g>0$; the edge of the spectrum in this case corresponds to a
homogeneous state with $\varphi =0$, the energy of this state is
$E=0$. We will start from the case of positive $g$; for negative
$g$ the Hamiltonian (\ref{hamm}) should be modified, to bound the
ground-state energy from below. This case will be considered
later, in Section 4.

In the mean-field approximation, at finite temperatures the band
edge is shifted up by
\begin{equation}
E_0 = g \langle \left| \nabla h\right|^2 \rangle \label{shift}
\end{equation}
The inclusion of fluctuations, however, changes this conclusion
dramatically. For the case of spin polarons, it was demonstrated
already by Brinkman and Rice \cite{brinkman} that the band
\textit{edges} do not depend on the degree of spin disorder. The
band is narrowed in antiferromagnetic or magnetically disordered
state in comparison with the ferromagnetic one but not due to the
shift of the band edges, just the density of states (DOS) becomes
larger at the middle of the band and exponentially small near the
edges, similar to the so called Lifshitz tails \cite{lifshitz} in
disordered systems. One can expect the same behavior for our case
(Fig. 1c). The mean-field energy (\ref{shift}) is, rather, an
inflection point of DOS; the states between $E=E_0$ and $E=0$ are
associated with rare events, namely, correlated fluctuations such
that in some large areas the membrane is flat and normals are
parallel. The tail of the DOS below $E_0$ is associated with
flexurons. The autolocalization region becomes larger when the
energy goes closer to the band edge $E=0$ (and the edge itself
corresponds to the completely flat state of the whole membrane,
cf. Refs. \onlinecite{brinkman,AK1}).

\begin{figure}
\begin{center}

\begin{minipage}{90mm}
\parbox{10mm}{(a)}\parbox{80mm}{\includegraphics[width=80mm]{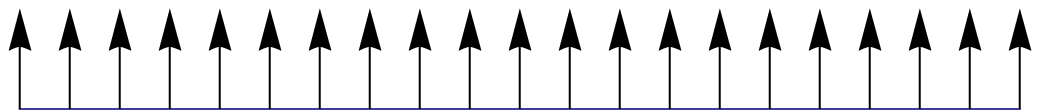}}
\end{minipage}

\begin{minipage}{90mm}
\parbox{10mm}{(b)}\parbox{80mm}{\includegraphics[width=80mm]{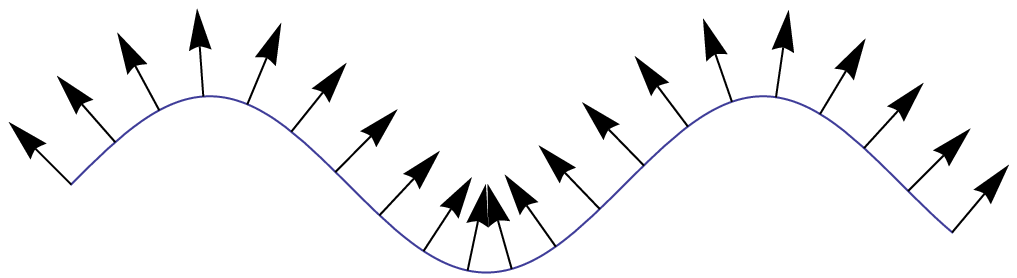}}
\end{minipage}

\begin{minipage}{90mm}
\parbox{10mm}{(c)}\parbox{80mm}{\includegraphics[width=80mm]{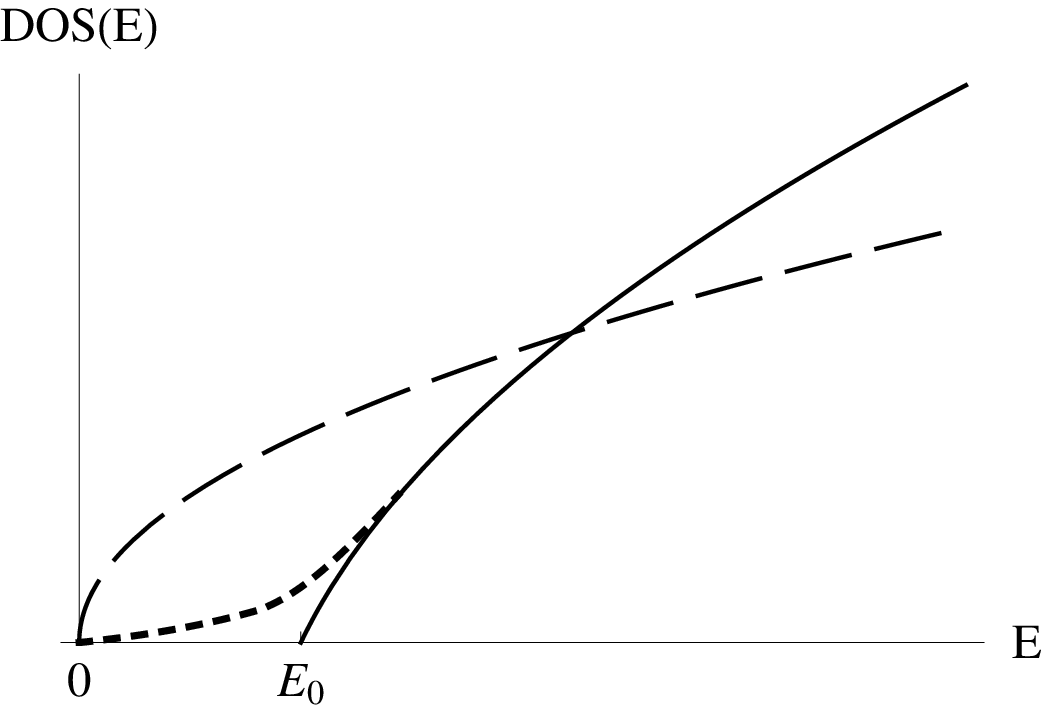}}
\end{minipage}
\end{center}
\caption{\label{fig::f1} Ground state of the membrane, all normals
(black arrows) are along the $z$ axis (a), and a snapshot of a
fluctuating state at finite temperatures (b). Panel (c) shows a
sketch of the density of states (DOS) at zero temperature (dashed
line), at finite temperatures within mean-field approximation
(solid line) and the fluctuation-induced tail (dotted line). We
assume $g>0$.}
\end{figure}

In crystalline membranes, the anharmonic effects are essential,
namely, the coupling between bending and stretching modes; without
this coupling the membrane turns out to be crumpled at any finite
temperatures \cite{nelson}. In the harmonic approximation which is
valid for not too small wavevectors $\mathbf{q}$ of fluctuations,
the Fourier component of the normal-normal correlation
function $G\left( \mathbf{q}\right) =q^2\left\langle \left| h_{\mathbf{q}%
}\right| ^2\right\rangle $ at temperature $T$ is given by the
expression \cite{nelson}
\begin{equation}
G\left( \mathbf{q}\right) =\frac T{\kappa q^2}  \label{harm}
\end{equation}
where $\kappa $ is the bending rigidity. The use of this
expression to calculate $\left\langle \left| \nabla h\right|
^2\right\rangle $ leads to a logarithmically divergent result. The
point is that the expression (\ref{harm}) is not applicable for
small enough $q$ where the anharmonic effects become dominant. The
coupling between the bending and the stretching phonons in
crystalline membranes leads to the renormalization of the bending
rigidity and its growth with $q$ decrease. Perturbation analysis
shows that the anharmonic effects become dominant at
\begin{equation}
q<q^{*}=\sqrt{3TY/8\pi \kappa ^2} \label{ginzb}
\end{equation}
where $Y$ is the Young modulus (``Ginzburg criterion''
\cite{nelson,Fas2007}). For smaller $q$,
\begin{equation}
G\left( \mathbf{q}\right) =\frac{A}{q^{2-\eta }q_0^\eta }
\label{anharm}
\end{equation}
where $q_0=\sqrt{Y/\kappa }$ is of the order of inverse
interatomic distance and $\eta $ is the exponent of the
renormalized bending rigidity. Within the self-consistent
screening approximation \cite{radz,gaz1,rafa} $\eta \simeq 0.821$.
Recent Monte Carlo simulations \cite{los}  and functional
renormalization group analysis \cite{kow,brag} give $\eta \simeq
0.85.$

The factor $A$ can be found from the matching of expressions
(\ref{harm}) and (\ref{anharm}) at $q=q^{*}$:
\begin{equation}
A = \alpha \left( \frac{T}{\kappa} \right)^{1-\eta/2} \label{A}
\end{equation}
where $\alpha$ is a numerical factor.

With logarithmic accuracy, one finds
\begin{equation}
E_0=\frac{Tg\Lambda }{2\pi \kappa }  \label{E00}
\end{equation}
where $\Lambda =\ln \left( q_0/q^{*}\right) \approx \left(
1/2\right) \ln \left( \kappa /T\right) $ originates from the
cutoff at small $q$ and we assume $T\ll \kappa $ (for example, for
graphene $\kappa \simeq 1eV$ \cite {Fas2007}). Thus, within the
mean-field approximation, the shift of the band edge is
proportional to temperature. Physically this effect is observable
(i.e. not hidden by thermal smearing) if $\left| E_0\right| >T,$ that is, $%
\left| g\right| \Lambda \gg \kappa $. Further we will assume that
this condition is fulfilled, e.g., due to a large logarithm.

\section{Fluctuation effects in the case of electron preferring flat surrounding}

To take into account the effect of fluctuations on the electron
energy spectrum I will use the path integral formalism
\cite{path1,path2,path3}, following
Refs.\onlinecite{ourTMF1,ourJMMM,AK1,AK2}.

The partition function of the whole system (the electron plus the
fluctuating field $\varphi$) is represented as
\begin{widetext}
\begin{equation}
Z=\mbox{Tr}e^{-\beta \mathcal{H}_{f}\left( \varphi \right) -\beta \mathcal{H}%
_{e}\left( \mathbf{r,}\varphi \right) }=Z_{f}\left\langle \mbox{Tr}_{\mathbf{%
r}}T_{\tau }\exp \left[ -\int_{0}^{\beta }\mathcal{H}_{e}\left(
\mathbf{r,}\varphi \left( \mathbf{r,}\tau \right) \right) d\tau
\right] \right\rangle _{f}  \label{partfun}
\end{equation}
\end{widetext}
where $\beta =T^{-1}$ is the inverse temperature, $\mathcal{H}_e$
is the Hamiltonian (\ref{hamm}), $Z_{f}=\mbox{Tr}_{\varphi
}e^{-\beta \mathcal{H}_{f}\left( \varphi
\right) }$ is the partition function of the field, $\mathcal{H}_{f}$ is the corresponding Hamiltonian, $\varphi \left( \mathbf{r,%
}\tau \right) =e^{\tau \mathcal{H}_{f}\left( \varphi \right)
}\varphi \left( \mathbf{r}\right) e^{-\tau \mathcal{H}_{f}\left(
\varphi \right) }$ and
\begin{equation}
\left\langle \mathcal{A}\left( \varphi \right) \right\rangle _{f}=\frac{1}{%
Z_{f}}\mbox{Tr}_{\varphi }e^{-\beta \mathcal{H}_{f}\left( \varphi \right) }%
\mathcal{A}\left( \varphi \right)   \label{meanf}
\end{equation}
is the average over the field states. Using the Feynman
path-integral approach \cite{path1,path2,path3,feynman} and taking
average over $\varphi$ yields for the electron-only free energy
\begin{equation}
\mathcal{F}=-\frac{1}{\beta }\left( \ln Z-\ln Z_{f}\right) =-\frac{1}{\beta }%
\ln \int_{\mathbf{r}\left( 0\right) =\mathbf{r}\left( \beta
\right) }e^{-\mathcal{S}}\mathcal{D}\left[ \mathbf{r}\left( \tau
\right) \right] , \label{freen}
\end{equation}
where the effective action of electron is
\begin{equation}
\mathcal{S}\left[ \mathbf{r}\left( \tau \right) \right]
=\mathcal{S}_0\left[ \mathbf{r}\left( \tau \right) \right]
+\mathcal{S}_{int}\left[ \mathbf{r}\left( \tau \right) \right]
\label{S1}
\end{equation}
and
\begin{equation}
\mathcal{S}_0\left[ \mathbf{r}\left( \tau \right) \right] =\frac
12\int\limits_0^\beta \left( \frac{d\mathbf{r}}{d\tau }\right)
^2d\tau,   \label{S2}
\end{equation}
\begin{equation}
\exp \left( -\mathcal{S}_{int}\left[ \mathbf{r}\left( \tau \right)
\right] \right)
=\left\langle \exp \left[ -g\int\limits_0^\beta \varphi \left[ \mathbf{r}%
\left( \tau \right) \right] d\tau \right] \right\rangle_f
\label{S3}
\end{equation}
(we put temporarily $\hbar =m=1$). The average can be formally
written as a cumulant expansion
\begin{equation}
\mathcal{S}_{int} =-\sum_{m=1}^{\infty
}\frac{g^{m}}{m!}\int_{0}^{\beta }...\int_{0}^{\beta
}\mathcal{K}_{m}\left( \mathbf{r}\left( \tau _{1}\right)
...\mathbf{r}\left( \tau _{m}\right) \right) d\tau _{1}...d\tau
_{m}
\end{equation}
where $\mathcal{K}_{m}\left(
\mathbf{r}_{1};...;\mathbf{r}_{m}\right) $ are the $m$-th cumulant
correlators, defined recursively by
\begin{align}
\mathcal{K}_{1}\left( \mathbf{r}_{1}\right) & =\left\langle \varphi \left(
\mathbf{r}_{1}\right) \right\rangle _{f},  \notag \\
\mathcal{K}_{2}\left( \mathbf{r}_{1};\mathbf{r}_{2}\right) & =\left\langle
\varphi \left( \mathbf{r}_{1}\right) \varphi \left( \mathbf{r}_{2}\right)
\right\rangle _{f}-\mathcal{K}_{1}\left( \mathbf{r}_{1}\right) \mathcal{K}%
_{1}\left( \mathbf{r}_{2}\right) ,...  \label{cumulant}
\end{align}%
etc.

To estimate the electron free energy $\mathcal{F}$ we use the same trial action as in Refs.\onlinecite%
{ourTMF1,ourJMMM,AK1},
$\mathcal{S}_{t}=\mathcal{S}_{0}+\mathcal{S}_{pot}$ where
\begin{equation}
\mathcal{S}_{pot}=\frac{\omega ^{2}}{4\beta }\int_{0}^{\beta
}\int_{0}^{\beta }\left[ \mathbf{r}\left( \tau \right) -\mathbf{r}\left(
\sigma \right) \right] ^{2}d\tau d\sigma ,  \label{tract}
\end{equation}%
the oscillator frequency $\omega $ being trial parameter (in
contrast with Ref.\onlinecite{AK2,feynman} we do not introduce any
retardation in the trial action since it is important only for the
case of {\it quantum} fluctuations of the field $\varphi$ which is
beyond the scope of the present work). It describes in a
translationally invariant way a self-trapped state if $\beta \hbar
\omega \gg 1$, in this case the size of the self-trapping region
is of the order of the zero-point oscillations of a harmonic
oscillator, $l = \sqrt{\hbar /2m \omega}$. The trial action
$\mathcal{S}_t$ gives an upper estimate of free energy for our
problem. The Peierls-Feynman-Bogoliubov inequality reads
\begin{equation}
\mathcal{F}\leq \mathcal{F}_{t}+\frac{1}{\beta }\left\langle \mathcal{S}%
_{int}-\mathcal{S}_{pot}\right\rangle _{t}  \label{PBineq}
\end{equation}
where $\mathcal{F}_{t}$ is the free energy corresponding to the
trial action $\mathcal{S}_{t}$, which is equivalent to
\begin{align}
\mathcal{F}& \leq \mathcal{F}_{t}-\frac{\omega ^{2}}{4\beta ^{2}}%
\int_{0}^{\beta }\int_{0}^{\beta }\left\langle \left[ \mathbf{r}\left( \tau
\right) -\mathbf{r}\left( \sigma \right) \right] ^{2}\right\rangle _{t}d\tau
d\sigma  \notag \\
& -\sum_{m=1}^{\infty }\frac{g^{m}}{m!\beta }\int_{0}^{\beta
}...\int_{0}^{\beta }\left\langle \mathcal{K}_{m}\left(
\mathbf{r}\left( \tau _{1}\right) ,...,\mathbf{r}\left( \tau
_{m}\right) \right) \right\rangle _{t}\prod_{j=1}^{m}d\tau _{j}
\label{interim1}
\end{align}

To proceed, we will pass to the Fourier transforms of the cumulants $%
\mathcal{K}_{m}\left( \mathbf{q}_{1},..,\mathbf{q}_{m-1}\right) $
and take into account that for the Gaussian trial action
$\mathcal{S}_{t}$ one has
\begin{align}
& \left\langle \exp \left\{ i\sum_{j=1}^{m-1}\mathbf{q}_{j}\left[ \mathbf{r}%
\left( \tau _{j}\right) -\mathbf{r}\left( \tau _{m}\right) \right] \right\}
\right\rangle _{t}=  \notag \\
& \exp \left\{ -\sum_{j,k=1}^{m-1}\frac{\mathbf{q}_{j}\mathbf{q}_{k}}{4}%
\left\langle \left[ \mathbf{r}\left( \tau _{j}\right) -\mathbf{r}\left( \tau
_{m}\right) \right] \left[ \mathbf{r}\left( \tau _{k}\right) -\mathbf{r}%
\left( \tau _{m}\right) \right] \right\rangle _{t}\right\} .
\label{interim3}
\end{align}%

For the autolocalized states the variational parameter $\omega $
satisfies the inequalities $\beta \omega \gg 1$ and $\omega \ll W$
where $W$ is of the order of the electron bandwidth (the last
inequality just means that the flexuron size $l=1/\sqrt{2\omega }$
is much larger than the interatomic distance, otherwise our
continuum-medium description is not applicable).

Due to the first inequality one has \cite{ourJMMM}
\begin{equation}
\left\langle \mathbf{r}\left( \tau _i\right) \mathbf{r}\left( \tau
_k\right) \right\rangle _t\approx 2l^2\delta _{ik}  \label{delta}
\end{equation}
and Eq.(\ref{interim3}) is simplified:
\begin{align}
& \mathcal{F}\leq \frac{\omega}{2} -\sum\limits_{m=1}^{\infty }\frac{1}{m!\beta}%
\left( g\beta \right) ^{m}\sum_{\mathbf{q}_{i}} \mathcal{K}_{m}\left( {\mathbf{q}}%
_{1},...,{\mathbf{q}}_{m-1}\right) \times   \notag \\
& \exp \left[ {-\frac{1}{{4}\omega }\sum\limits_{j=1}^{m-1}{\mathbf{q}_{j}^{2}}%
-\frac{1}{{4}\omega }\left(
{\sum\limits_{j=1}^{m-1}{\mathbf{q}_{j}}}\right) ^{2}}\right].
\label{F}
\end{align}%

In harmonic approximation, the second cumulant can be calculated
using Wick's theorem:
\begin{equation}
\mathcal{K}_2\left( \mathbf{q}\right)
=\sum\limits_{\mathbf{p}}G\left( \mathbf{q}\right) G\left(
\mathbf{p-q}\right).   \label{wick}
\end{equation}
Within the self-consistent screening approximation \cite{radz}
(neglecting vertex corrections) this expression is also supposed
to be correct. Note that this approximation is in reasonable
agreement with the results of
Monte Carlo simulations \cite{rafa}. We will use it here. Substituting Eqs.(%
\ref{harm}) and (\ref{anharm}) into Eq.(\ref{wick}) one finds
\begin{equation}
\mathcal{K}_2\left( \mathbf{q}\right) =\left\{
\begin{array}{cc}
\frac 1\pi \left( \frac T\kappa \right) ^2\frac{\ln \left( q/q^{*}\right) }{%
q^2}, & q>q^{*} \\
\frac{\alpha ^{\prime }}{q_0^{2\eta }q^{2-2\eta }}\left( \frac
T\kappa \right) ^{2-\eta }, & q<q^{*}
\end{array}
\right.   \label{cc}
\end{equation}
where $\alpha ^{\prime }$ is a numerical factor.

Let us start estimating Eq.(\ref{F}) in the Gaussian
approximation, that is, taking into account only terms with
$m=1,2$, then
\begin{equation}
\mathcal{F}\leq E_0+\frac 1{4l^2}-\frac{g^2}{2T}\sum\limits_{\mathbf{q}}%
\mathcal{K}_2\left( \mathbf{q}\right) \exp \left( -q^2l^2\right)
\label{gauss}
\end{equation}
Now we have to minimize the right-hand-side of Eq.(\ref{gauss}) to
find the
size of flexuron $l_0$ and the autolocalization energy.

Let us assume first that $q^{*}l\ll 1$. In this case one can
rewrite Eq.(\ref {gauss}) as
\begin{equation}
\mathcal{F}\leq E_0+\frac{3TY\hbar^2}{32\pi \kappa
^2m}x-\frac{Tg^2}{32\pi ^2\kappa ^2}\ln ^2x,  \label{gauss1}
\end{equation}
where $x=\left( q^{*}l\right) ^{-2}$ and we have restored the
missing constants $\hbar$ and $m$.

The last two terms in the right-hand-side of Eq.(%
\ref{gauss1}) are proportional to temperature. The minimum of this
function exists if
\begin{equation}
\nu =\frac{g^2m}{3\pi Y\hbar ^2}>\frac e2  \label{nu}
\end{equation}
($e\simeq 2.71828...$) which is a criterion of autolocalization in
this approximation. Of course it is valid only with an accuracy of
some numerical factor since we kept only leading logarithms in our
estimations. It is hardly to expect that $\nu $ is {\it much}
larger than one, thus, an optimal value of $x$ is
of the order of unity, beyond the formal limit of the approximation (\ref{gauss1}%
). The analysis from the opposite limit, $q^{*}l\gg 1$ gives the
same result which is not surprising keeping in mind that the
expressions (\ref{cc}) match at $q \approx q^*$. Therefore,
\begin{eqnarray}
l_0 &\simeq &1/q^{*}\propto T^{-1/2},  \nonumber \\
\mathcal{F}_{opt} &\simeq &E_0-const \cdot T  \label{opt}
\end{eqnarray}
This is the main result of our consideration, and it is amazingly
simple: the size of the self-trapping region is of the order of
$1/q^{*}$ (which is much larger than the interatomic distance, so
the continuum medium description works). Thus, inhomogeneities
induced by electrons in fluctuating membrane should have the same
spatial scale as the crossover length from the harmonic to the
anharmonic regime.

The higher-order terms in the cumulant expansion in Eq.(\ref{F})
can be estimated from the expressions in the harmonic regime. Each
next couple of the Green functions $G\left( \mathbf{q}\right) $
gives an additional factor $1/q^2$, together with additional
integration over $\mathbf{q}$ (which can affect the powers of
logarithms) and, importantly, the additional small parameter
$\left( T/\kappa \right) ^2$, which is only partially compensated
by the additional factor $\beta g$. Thus, the cumulant expansion
has a small parameter $T g/\kappa^2$ and the results of the
Gaussian approximation are reliable.

A compact analytical expression for the density of states $N(E)$
can be obtained for the Lifshitz tail regime, that is, very close
to the true edge of the spectrum $E=0$. To this aim, one can use
the method of Ref.\onlinecite{AK1}. For the range of energies $E
\gg T$, relevant flat regions have a size much larger than
$1/q^{*}$, thus, we are in the strongly anharmonic region. From
scaling considerations \cite{PP}, one can postulate that for all
$m$, as well as for $m=2$ (cf. Eq.(\ref{cc})),
\begin{equation}
\mathcal{K}_{m}\left(
{\mathbf{q}}_{1},...,{\mathbf{q}}_{m-1}\right)
=a^{\left( 2-2\eta \right) \left( m-1\right) }\mathcal{K}_{m}\left( a{\mathbf{%
q}}_{1},...,a{\mathbf{q}}_{m-1}\right).
\end{equation}
Further consideration just repeats that of Ref.\onlinecite{AK1}
for the case of XY-model. The result for the Lifshitz tail of the
density of states is
\begin{equation}
\ln N\left( E\right) \propto -\left( \frac{T}{E }\right)
^{\eta}\left( \ln \frac{T}{E} \right)^{\frac{1}{2\eta+1}},
\label{tail}
\end{equation}
($E \ll T$). Of course, probing experimentally the states with
energies much smaller than the temperature is not an easy task, so
the expression (\ref{tail}) is mainly of formal interest.

The number of electron states within the tail between $E=0$ and
$E=E_0$ can be estimated by a diagrammatic approach suggested in
Ref.\onlinecite{AK2}, Section 4.3. By analogy with Eq.(152) of
that work, one can estimate the electron concentration per atom as
\begin{equation}
n_c \approx gT/\kappa^2. \label{cap}
\end{equation}

\section{Fluctuation effects in the case of electrons preferring crumpled surrounding}

Consider now the case of a negative coupling constant which
physically means that corrugations result in a gain of electron
energy. The Hamiltonian (\ref {hamm}) cannot be used here since
its spectrum is not bounded from below. Instead we will use the
Hamiltonian
\begin{equation}
H=-\frac{\hbar ^2}{2m}\nabla ^2+\frac g2\left( \varphi \left( \mathbf{r}%
\right) -\varphi _0\right) ^2  \label{hamm1}
\end{equation}
where $g$ is still positive and $\varphi _0$ is the value of
$\left| \nabla
h\right| ^2$ optimal for electron hopping; the coupling constant is now $%
g_0=-g\varphi _0$, and $E=0$ is still the lowest possible energy.
Using the Hubbard-Stratonovich transformation, the partition
function of electron in a given field $\varphi \left(
\mathbf{r}\right) $ can be represented as (again, $\hbar =m=1$):
\begin{widetext}
\begin{eqnarray}
Z_e &=&\int \mathcal{D}\mathbf{r}\left( \tau \right) \exp \left[
-\frac
12\int\limits_0^\beta \left( \frac{d\mathbf{r}\left( \tau \right) }{d\tau }%
\right) ^2d\tau -\frac g2\int\limits_0^\beta \left[ \varphi \left( \mathbf{r}%
\left( \tau \right) \right) -\varphi _0\right] ^2d\tau \right] =
\nonumber
\\
&&\int \mathcal{D\gamma }\left( \tau \right) \int \mathcal{D}\mathbf{r}%
\left( \tau \right) \exp \left[ -\frac 12\int\limits_0^\beta \left( \frac{d%
\mathbf{r}\left( \tau \right) }{d\tau }\right) ^2d\tau
-i\int\limits_0^\beta \mathcal{\gamma }\left( \tau \right) \left[
\varphi \left( \mathbf{r}\left( \tau \right) \right) -\varphi
_0\right] d \tau -\frac 1{2g}\int\limits_0^\beta \mathcal{\gamma
}^2\left( \tau \right) d\tau \right]  \label{hs}
\end{eqnarray}
\end{widetext}
Repeating the derivation of Eq.(\ref{interim1}) we will find the
only difference: $g^m$ is replaced by the product
$i^m\prod_{j=1}^{m}\gamma \left( \tau _{j} \right)$ averaged over
fluctuations of the Gaussian field $\gamma (\tau)$.

Continuing the transformations leading from Eq.(\ref{interim1}) to
Eq.(\ref{F}) one obtains formally the same equation but with the
replacement of $g$ by $i\gamma_0$,
\begin{equation}
\gamma_0 = \frac{1}{\beta} \int\limits_0^\beta \gamma(\tau) d
\tau.
\end{equation}
At the end, as follows from Eq.(\ref{hs}), one has to average the
result over the Gaussian random {\it static} field $\gamma_0$
distributed with the probability function
\begin{equation}
P\left( \gamma _0\right) =\sqrt{\frac{\pi \beta }{2g}}\exp \left( -\frac{%
\beta \gamma _0^2}{2g}+i\beta \gamma _0\varphi _0\right) \label{P}
\end{equation}
An optimal value of this random field corresponding to the
extremum of the exponent in Eq.(\ref{P}) is equal to $ig\varphi_0$
and the fluctuations are of the order of $\sqrt{gT}$, that is,
negligible if $T \ll g\varphi_0$. Thus, the effective coupling
constant for the fluctuation contributions is just equal to $g_0$.
All the conclusions of the previous section about the size of
flexuron and criterion of autolocalization remain valid. The
mean-field expression for the shift of the band edge is now
different. As follows directly from Eq.(\ref{hamm1})
\begin{equation}
E_0 = \frac{g}{2} \left( \varphi_0^2 - \langle \varphi^2 \rangle
\right)^2.
\end{equation}

\section{Discussion and conclusions}

We have shown that the electron can be self-trapped by bending
fluctuations assuming that the parameter $\nu$, Eq.(\ref{nu}) is
large enough. One can expect $g$ to be of the order of the
electron bandwidth $W$ (cf. the discussions for graphene reviewed
in Ref.\onlinecite{VKG}). For the Young modulus a natural
estimation is $Y \approx E_{coh}/a^2$ where $E_{coh}$ is a
cohesive energy and $a$ is a lattice constant. For covalently
bonded membranes (such as graphene or h-BN) $W \approx E_{coh}$.
This means that the effective mass should be of the order of the
free electron mass, and therefore it is hardly to expect the
formation of flexuron for, e.g., gapped graphene with a mass $ m
\ll \hbar^2/ Wa^2$. On the other hand, the criterion of
autolocalization should be easy to fulfill for a soft membrane
with $E_{coh} \ll W$.

Assuming that we are in the regime of the autolocalization, the
electron injection into crystalline membranes at finite
temperatures does not result in the conductivity, the states will
be (auto)localized, even without external disorder. This will be
the case till the whole fluctuation tail will be occupied. The
number of the states in the fluctuation tail is given by
Eq.(\ref{cap}), so the transition to the conducting states happens
at $n = n_c \propto T$. The higher the temperature the more
electrons should be injected.

Strictly speaking, the flexuron states are completely localized
only if one assumes that the fluctuations are static. In an ideal
system, the flexuron can move together with the anomalously
corrugated region, but very slowly, for typical phonon times (cf.
the discussion for the case of fluctuon \cite{krivoglaz}). One can
expect, however, that even very weak extrinsic disorder will
localize so slow particle.

If the flexuron forms, its size is of the order of $1/q^*$ where
$q^*$ (\ref{ginzb}) is a crossover point from harmonic to
anharmonic regime, so, the bending fluctuations are stabilized by
electrons just at the border of the strong anharmonicity.

A very interesting issue for future studies is the case of finite
electron (flexuron) concentrations. By analogy with magnetic
semiconductors, one can expect a tendency to the phase separation
and electron droplet formation \cite{ourTMF2,nagaev,dagotto}.
Electronic mechanisms of stabilization of corrugations were
discussed in a context of graphene \cite{castroneto,gazit,guinea}
but in a broad-gap semiconductors considered here physics should
be essentially different. Probably, the flexuron formation is a
proper term in this case.

\section*{Acknowledgement}

I am thankful to C. Stampfer, A. Fasolino and A. Geim for
stimulating discussions. This work is part of the research program
of the Stichting voor Fundamenteel Onderzoek der Materie (FOM),
which is financially supported by the Nederlandse Organisatie voor
Wetenschappelijk Onderzoek (NWO).

\end{document}